%% file: main.tex
\documentclass[runningheads]{llncs}
\usepackage[utf8]{inputenc}  % Hỗ trợ UTF-8
\usepackage[T5]{fontenc}     % Hỗ trợ font tiếng Việt (nếu cần)
\usepackage{algorithm}
\usepackage{algorithmic}
\usepackage{svg}
\svgpath{{figs/}}
\usepackage{amssymb}
\usepackage{amsmath}
\usepackage{subfig}
\usepackage{hyperref}
\definecolor{pp-pink}{RGB}{221, 72, 151}

\usepackage{orcidlink}

\usepackage{graphicx}
\usepackage{hyperref}  % Thêm cái này để enable hyperlink cho citations và bibliography

\usepackage{color}
\hypersetup{
    colorlinks=true,
    linkcolor=blue,
    filecolor=blue,      
    urlcolor=blue,
    pdftitle={Overleaf Example},
    pdfpagemode=FullScreen,
    }
\renewcommand{\figurename}{Figure}
\begin{document}
\title{Enhanced Multimodal Video Retrieval System: Integrating Query Expansion and Cross-modal Temporal Event Retrieval}
%
%\titlerunning{Abbreviated paper title}
% If the paper title is too long for the running head, you can set
% an abbreviated paper title here

\author{
Van-Thinh Vo$^{1}$\orcidlink{0009-0009-4815-1397} \and
Minh-Khoi Nguyen$^{1}$\orcidlink{0009-0000-0813-6576} \and
Minh-Huy Tran$^{1}$\orcidlink{0009-0004-0535-5879} \and
Anh-Quan Nguyen-Tran$^{1}$\orcidlink{0009-0000-3966-1795} \and
Duy-Tan Nguyen$^{1}$\orcidlink{0009-0008-8732-7079} \and
Loi Nguyen Khanh$^{\dag}$\orcidlink{0000-0003-4300-491X} \and
Anh-Minh Phan$^{\dag}$\orcidlink{0009-0007-0217-5807}
}

%\author{Anonymous submission} 

\authorrunning{V.T. Vo et al.}
% First names are abbreviated in the running head.
% If there are more than two authors, 'et al.' is used.
%
\institute{Ho Chi Minh University of Technology, VNU-HCM, Vietnam, \\[2pt]
\texttt{\{thinh.vovan, khoi.nguyenminh82005, huy.tranminh2005, quan.nguyenquan12, tan.nguyen2313054, nkloi\}@hcmut.edu.vn, phananhm5@gmail.com}}

\title{Enhanced Multimodal Video Retrieval System: Integrating Query Expansion and Cross-modal Temporal Event Retrieval}
\titlerunning{Advanced Multimodal Video Retrieval System}
\maketitle % typeset the header of the contribution
\footnotetext[1]{These authors contributed equally to this work as first authors.}
\footnotetext[2]{$\dag$ Corresponding authors.}
\footnotetext[3]{ Our project page is available at \href{https://huyrua27.github.io/eeiotnewbie-AIC2025}{Project-Page}}
\begin{abstract}
Multimedia information retrieval from videos remains a challenging problem. While recent systems have advanced multimodal search through semantic, object, and OCR queries - and can retrieve temporally consecutive scenes - they often rely on a single query modality for an entire sequence, limiting robustness in complex temporal contexts. To overcome this, we propose a cross-modal temporal event retrieval framework that enables different query modalities to describe distinct scenes within a sequence. To determine decision thresholds for scene transition and slide change adaptively, we build Kernel Density Gaussian Mixture Thresholding (KDE-GMM) algorithm, ensuring optimal keyframe selection. These extracted keyframes act as compact, high-quality visual exemplars that retain each segment's semantic essence, improving retrieval precision and efficiency. Additionally, the system incorporates a large language model (LLM) to refine and expand user queries, enhancing overall retrieval performance. The proposed system's effectiveness and robustness were demonstrated through its strong results in the Ho Chi Minh AI Challenge 2025.

\end{abstract}
\input{contents/Intro}
\input{contents/RelatedWork}
\input{contents/method}
\input{contents/UI}

\input{contents/conclusion}
\section{Acknowledgement}
We acknowledge Ho Chi Minh City University of Technology (HCMUT), VNU-HCM, for supporting this study.

%
% ---- Bibliography ----
%
% BibTeX users should specify bibliography style 'splncs04'.
% References will then be sorted and formatted in the correct style.
%
% \bibliographystyle{splncs04}
% \bibliography{mybibliography}
%
%\begin{thebibliography}{8}
%\bibitem{ref_article1}
%Author, F.: Article title. Journal \textbf{2}(5), 99--110 (2016)

%\bibitem{ref_lncs1}
%Author, F., Author, S.: Title of a proceedings paper. In: Editor,
%F., Editor, S. (eds.) CONFERENCE 2016, LNCS, vol. 9999, pp. 1--13.
%Springer, Heidelberg (2016). \doi{10.10007/1234567890}

%\bibitem{ref_book1}
%Author, F., Author, S., Author, T.: Book title. 2nd edn. Publisher,
%Location (1999)

%\bibitem{ref_proc1}
%Author, A.-B.: Contribution title. In: 9th International Proceedings
%on Proceedings, pp. 1--2. Publisher, Location (2010)

%\bibitem{ref_url1}
%LNCS Homepage, \url{http://www.springer.com/lncs}, last accessed 2023/10/25
%\end{thebibliography}

\bibliographystyle{splncs04}   
\bibliography{Ref}    

\end{document}

%% file: contents/Intro.tex
\section{Introduction} 
The exponential growth of personal video data - from lifelogs to surveillance footage - has created an urgent need for  efficient retrieval systems. Traditional keyword-based search proves inadequate when users must navigate hours of footage to locate specific moments, particularly when those moments lack textual metadata or predefined tags. Modern multimodal retrieval systems address this challenge by accepting diverse query modalities: natural language descriptions, visual examples, and metadata such as objects, text, etc. Recent advances in vision-language models, particularly CLIP \cite{laion} and BEiT3 \cite{wang2023image}, have demonstrated strong zero-shot capabilities by learning unified representations across modalities. Complementary techniques, including scene boundary detection~\cite{soucek2024transnet}, object detection~\cite{varghese2024yolov8}, and optical character recognition~\cite{smith2007overview,du2020pp} provide additional contextual signals for video understanding.
\\
Early video retrieval systems mainly targeted single-frame or isolated moment queries, but real-world applications now demand more advanced temporal reasoning. This evolution stems from the inherently temporal nature of human activities and the need to interpret video content as continuous narratives rather than independent frames. The AI Challenge Ho Chi Minh City 2025 \cite{AIChallenge2025} illustrates this progression: while the 2024 edition involved two-scene queries, the 2025 challenge doubles the complexity to four-scene temporal sequences, requiring deeper cross-frame reasoning and more robust temporal coherence modeling.

In this paper, we propose a comprehensive multimodal retrieval framework that addresses these challenges through several key innovations:

\begin{itemize}
\item \textbf{Adaptive Scene Boundary Detection:} We introduce a novel thresholding algorithm dynamically determines optimal cut-point thresholds for scene segmentation in TransNetV2~\cite{soucek2024transnet} and similar shot boundary detection networks, improving on traditional fixed thresholds.

\item \textbf{Query Enhancement via Large Language Models:} Integration of Gemini allows for intelligent query expansion, transforming natural language queries into enriched search expressions.

\item \textbf{Cross-modal Temporal Event Retrieval:} A temporal reasoning module identifies sequential event patterns across frames and modalities, accurately localizing complex actions by aligning multimodal representations over time.

% \item \textbf{Interactive User Interface with Temporal Navigation:} To facilitate efficient exploration of retrieval results, we design an interactive interface that presents search outcomes with direct timestamp linkages to source videos, enabling rapid verification and refinement of results.
\end{itemize}

Challenge results demonstrated the effectiveness of our approach on real-world video retrieval tasks, particularly in handling complex temporal queries that characterize modern interactive video search scenarios. Our team name in this challenge is EEIoT\_newbie.

%% file: contents/RelatedWork.tex
\section{Related Work}
\subsection{Importance and Applications of Video Retrieval}
% Video event retrieval has evolved significantly from traditional text-based methods to modern multimodal systems. Early approaches relied heavily on metadata such as titles, descriptions, and tags. However, these methods often failed when metadata was incomplete or ambiguous, particularly for complex or visually rich events that are difficult to describe with simple keywords. To address these challenges, recent research has explored more interactive and multimodal retrieval systems.

Video retrieval has become an essential research topic due to the rapid growth of multimedia content across online platforms. With the explosion of user-generated videos, efficient retrieval systems are crucial for organizing, accessing, and understanding large-scale video collections~\cite{rui1999image}. Applications of video retrieval span various domains, including surveillance analysis, content recommendation, video summarization, and cross-modal search in multimedia databases.
\subsection{Existing Systems and Innovations}
Currently, numerous systems have been developed to effectively address this challenge. To achieve robust and efficient free-text and semantic similarity retrieval, VISIONE~\cite{visione2021} integrates three powerful cross-modal feature extractors, each based on a pre-trained model. Additionally, it integrates a metadata-based search mechanism for object detection and object tagging, thereby enhancing the system’s capability for detailed information description.

At VBS 2024, query expansion techniques were a notable innovation to further improve retrieval precision~\cite{ma2024leveraging}. They applied large language models (LLMs) such as GPT-4 to rephrase and summarize user queries, enabling the system to capture key semantic information better. The expanded versions of queries were displayed to users, allowing them to select the most suitable one.
\subsection{Challenges in Temporal Queries and Proposed Strategy}
Although modern retrieval systems have made notable progress, handling temporally related queries in benchmark competitions remains a major challenge. Existing temporal reasoning methods perform well~\cite{visione2021,vuong2024viewsinsight}, but most depend on a single modality per scene, limiting their ability to model diverse and complex temporal dependencies. To overcome this, our work introduces a temporal retrieval strategy that integrates multiple modalities within a unified framework. By exploiting the complementary strengths of these heterogeneous channels, the system achieves stronger temporal reasoning, improving both robustness and precision in complex video retrieval scenarios.

%% file: contents/method.tex
\section{Method}
\subsection{Data Preprocessing}
\begin{figure}[ht]
    \centering
    \includegraphics[width=0.8\linewidth]{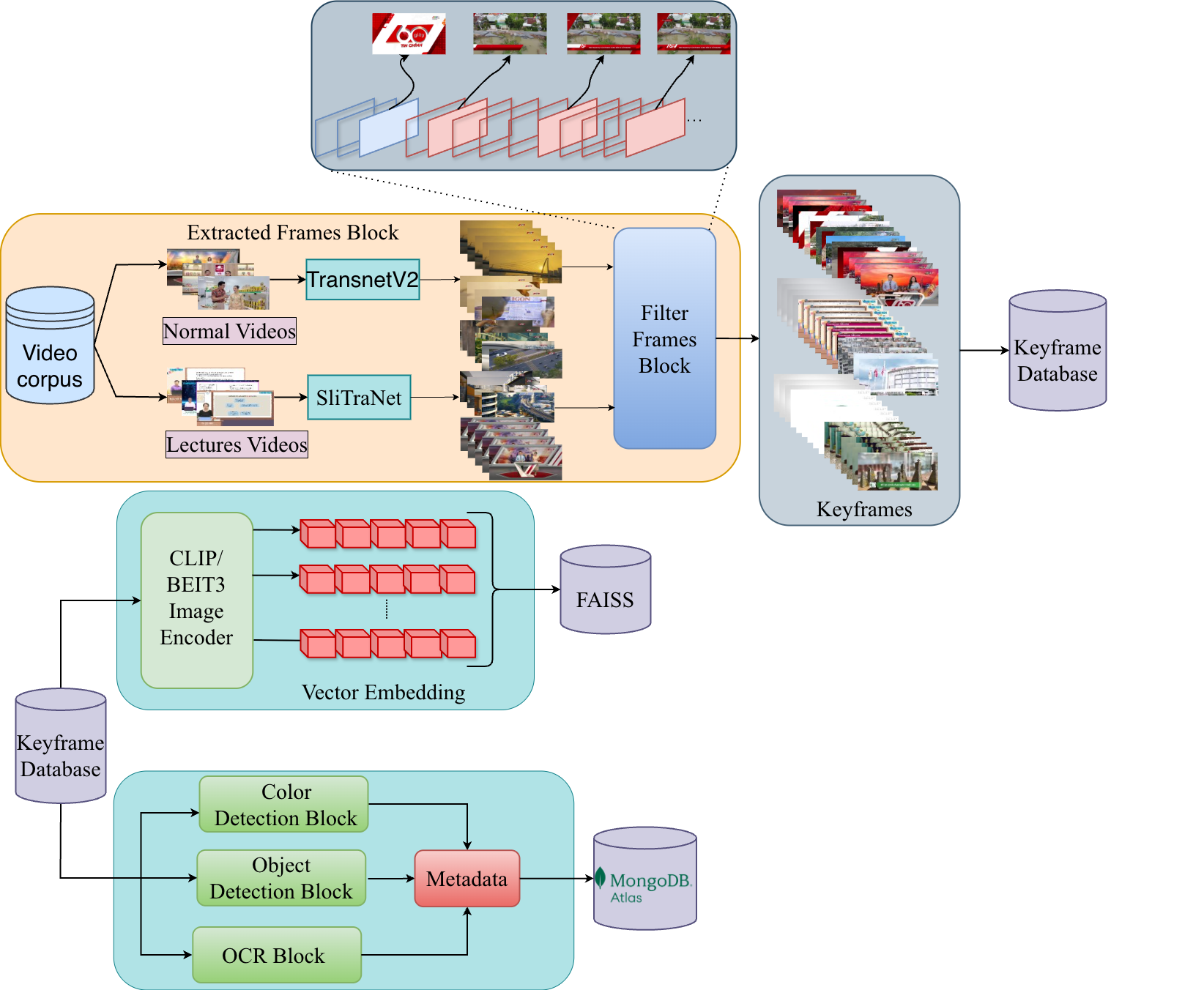}
\caption{Overview of the data preprocessing pipeline. Raw videos are first processed to extract representative keyframes stored in the Keyframe Database. Each keyframe is then analyzed through BEiT3/CLIP feature extraction, OCR, object detection, and color detection modules. The outputs are consolidated into structured metadata for downstream multimodal retrieval.}
    \label{fig:pre}
\end{figure}

\subsubsection{Frame Extraction:} 

To determine the optimal binary classification threshold $\theta^*$ for models like TransNetV2 and SliTraNet\cite{sindel2022slitranet}, we propose the Kernel Density Gaussian Mixture Thresholding algorithm (\hyperref[alg:kde_gmm]{Algorithm 1}). This method minimizes the Bayes error by approximating the score distribution via Kernel Density Estimation (KDE)~\cite{parzen1962estimation} to initialize a two-component Gaussian Mixture Model (GMM). The GMM parameters are refined using the Expectation Maximization (EM) algorithm~\cite{mclachlan2019finite}, determining $\theta^*$ at the intersection of the two components.

Subsequently, to extract representative frames (\hyperref[fig:pre]{Figure 1}), we apply K-Means ($K=3$)\cite{ahmed2020k} on deep features extracted via MobileNetV2\cite{sandler2018mobilenetv2}. The frames closest to the cluster centroids are selected as high-quality exemplars for downstream processing.
\begin{algorithm}[ht]
    \caption{Kernel Density and Gaussian Mixture Thresholding}
    \label{alg:kde_gmm}
        \begin{algorithmic}[1]
            \REQUIRE Scores $\{p_i\}_{i=1}^N$, bandwidth $h$, kernel $K(\cdot)$
            \ENSURE Estimated threshold $\theta^*$
            \STATE \textbf{KDE step:} Estimate overall density 
            $\hat f(x)=\frac{1}{Nh}\sum_i K\!\left(\frac{x-p_i}{h}\right)$.
            \STATE \textbf{Find modes:} Detect two main peaks 
            $m_1<m_2$ and local minimum $b$ between them.
            \STATE \textbf{Region split:} 
            $\mathcal{C}_1=\{p_i\!\le b\}$,\;
            $\mathcal{C}_2=\{p_i\!> b\}$.
            \STATE \textbf{Initial parameters:}
            $\hat\pi_k=\tfrac{|\mathcal{C}_k|}{N}$,\;
            $\hat\mu_k,\hat\sigma_k$ from region samples with $k\in\{1,2\}$
            \STATE \textbf{GMM refinement (EM):} 
            Run EM updates:
            $
            \gamma_{ik}=\frac{w_k\mathcal{N}(p_i|\mu_k,\sigma_k^2)}
            {\sum_j w_j\mathcal{N}(p_i|\mu_j,\sigma_j^2)},
            $
            update
            $w_k,\mu_k,\sigma_k^2$ until convergence.
            \STATE \textbf{Threshold solving:} 
            Find all intersections 
            $\mathcal{X}=\{x\in[\min(\mu_1,\mu_2),\max(\mu_1,\mu_2)] : w_1\phi_1(x)=w_2\phi_2(x)\}$.
            If $\mathcal{X}\ne\varnothing$, set $\theta^*=\arg\min_{x\in\mathcal{X}} E[\mathrm{error}(x)]$; otherwise set $\theta^*=\arg\min_{t\in\{\min(\mu_1,\mu_2),\max(\mu_1,\mu_2)\}} E[\mathrm{error}(t)]$.
            \RETURN $\theta^*$
        \end{algorithmic}
\end{algorithm}

\subsubsection{Visual Feature Extraction:}
For visual representation, we employ pretrained vision-language models CLIP and BEiT3 as feature extractors. The extracted features are saved into the FAISS \cite{douze2024faiss} index and subsequently used for retrieval and alignment with text representations in our framework.

\subsubsection{Metadata Representation:} To enable diverse and efficient search capabilities, our system stores rich metadata extracted from keyframes into MongoDB Atlas database. Each keyframe is associated with several metadata components, including:
\begin{figure}
        \centering
        \includegraphics[width=0.9\linewidth]{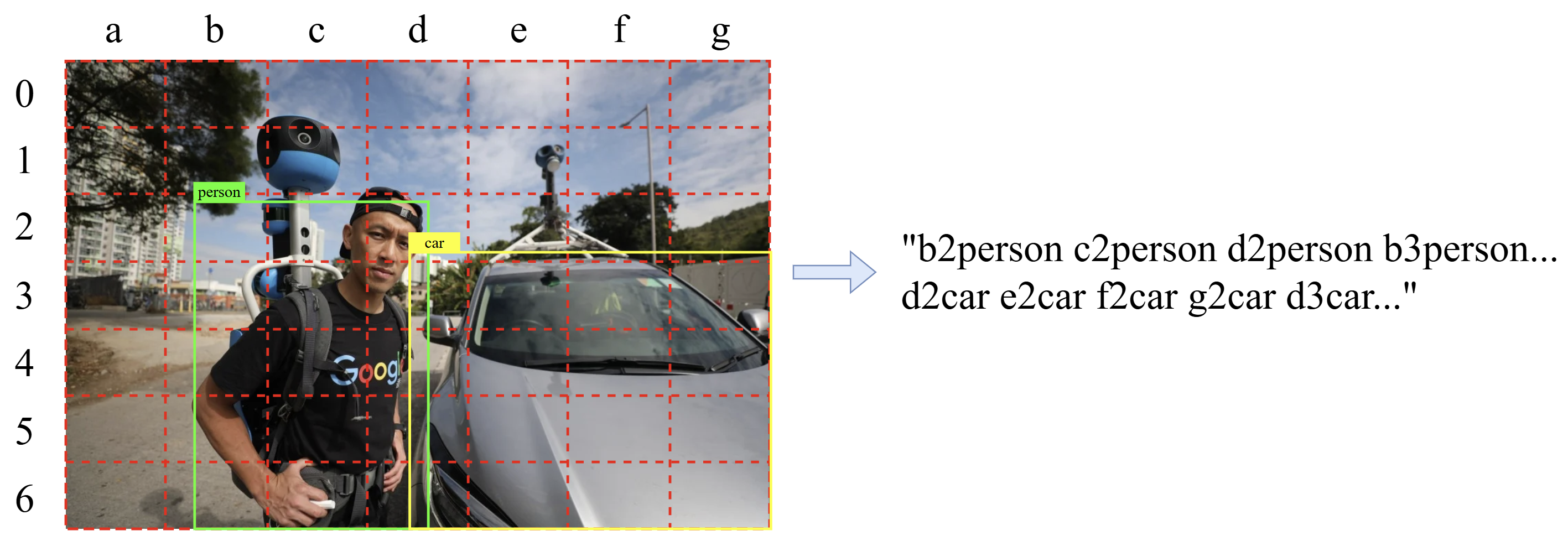}
        \caption{After detecting objects in the frame, we encode them into a string of text-based bounding boxs as above. In addition, the other metadata collected are the object tags and counts, e.g. "person1 car1".}
        \label{fig:canvas}
    \end{figure}
\begin{itemize}
    \item \textbf{Object Detection}: We used two YOLO versions for object detection: YOLOv11 \cite{yolo11} pretrained on the COCO dataset~\cite{coco} and YOLOv8~\cite{yolo8} pretrained on the OpenImagesV7 dataset. Both models apply a 7x7 grid over the image, with each cell representing a defined region. Object class and position are encoded using codloc (location code) and codclass (class code), forming a textual representation of object distribution. See \hyperref[fig:canvas]{Figure 2} for more details.

    \item \textbf{Color Detection}: The study limits the color palette to 11 universal color terms (white, black, red, green, yellow, blue, brown, purple, pink, orange, and gray). To implement this, two chip-based color naming methods~\cite{benavente2008parametric,van2009learning} were used, employing Probabilistic Latent Semantic Analysis and a parametric fuzzy model to map RGB values to color names. These models label pixels with color terms and determine dominant colors for 7x7 image cells.
    
    \item \textbf{Optical Character Recognition}: The OCR module extracts text from video keyframes using a multi-stage pipeline. First, the CRAFT model~\cite{baek2019character} detects and localizes text regions, followed by the ViT-B/16 Encoder~\cite{dosovitskiy2020image}, which classifies text as handwritten, printed, or mixed. Depending on the type, the system applies EasyOCR~\cite{salehudin2023analysis} and TesseractOCR~\cite{smith2007overview} for printed text, PaddleOCR~\cite{cui2025paddleocr} for typewritten text, and SFR~\cite{wigington2018start} for handwritten or artistic text. Finally, LayoutLMv3~\cite{xu2020layoutlm} refines results using contextual and spatial cues, improving accuracy and consistency. The processed text is then stored and indexed for efficient video retrieval.
\end{itemize}

\subsection{System Overall}
\begin{figure}[ht]
    \centering
    \includegraphics[width=0.9\linewidth]{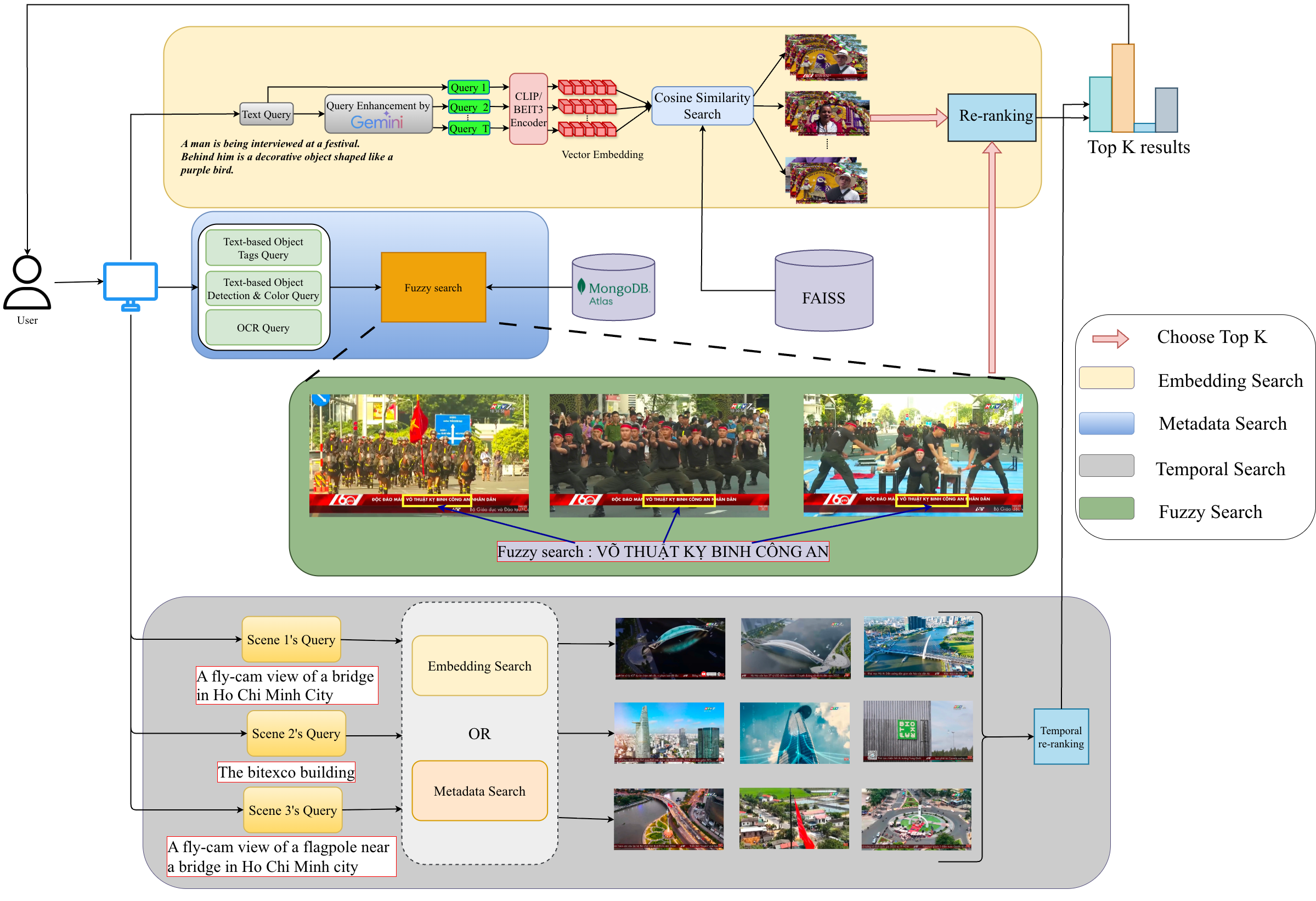}
    \caption{The Overview of Retrieval System. We propose two main search engines: Embedding-based Search and Metadata-based Search. To enhance the system performance, we also integrate them to address multimodal query and temporal events retrieval.}
    \label{fig:search}
\end{figure}
\subsubsection{Embedding-based Search:}

We use two vision-language encoders, CLIP and BEiT3, to encode each textual query into an embedding. The query embedding is compared with precomputed frame embeddings in a FAISS~\cite{douze2024faiss} index using cosine similarity to return the top-k most relevant keyframes. When Gemini-based~\cite{comanici2025gemini} is used for query augmentation of $q$, it generates a set of queries $Q=\{q_1, q_2, q_3, ..., q_n\}$ that are semantically equivalent to the original query. The system employs Google’s Gemini API, thereby mitigating the need for resource-intensive deployment. However, this improvement in query performance comes at the cost of increased response latency and restricted usage quotas. The system retrieves the top-k results for each of the $N$ augmented queries and then selects the top-k distinct keyframes with the highest similarity scores as the final results for $q$.

\subsubsection{Metadata-based Search:}
This component uses Fuzzy Search, an API from MongoDB Atlas, to retrieve objects whose textual attributes closely match the user query. Additionally, users can define logical relations (OR or AND) to refine search behavior across multiple bounding boxes, enabling flexible control over object-level semantic constraints.
\subsubsection{Re-ranking:}
To integrate the retrieval results obtained from multiple query type - including embedding-based search and metadata-based search - and to produce a unified ranking that reflects their overall relevance, we apply the Reciprocal Rank Fusion (RRF) method. RRF combines ranked lists by assigning each result a fusion score based on its position within individual rankings, as defined by the following formula:
\[
\mathrm{RRF}(d) = \sum_{i=1}^{n} \frac{1}{k + r_i(d)}
\] where $r_i(d)$ denotes the rank position of document (in our scenario, this is keyframe) $d$ in the i-th result list, and $k$ is a constant that controls the influence of lower-ranked items ($60$ by default). The fused list is then sorted according to the RRF scores, yielding a final ranking that best represents the collective relevance across all query modalities.

\subsubsection{Multimodal Search:}
To enhance retrieval performance, we introduce a multimodal query mechanism that lets users express search intent through multiple modalities simultaneously. Each modality is processed independently, and results are re-ranked using the Reciprocal Rank Fusion (RRF) method. As shown in \hyperref[fig:search]{Figure 3}, the Re-ranking block combines Embedding-based and Metadata-based Search results to produce the final top-k outputs.
\begin{algorithm}
\caption{Cross-modal Temporal Event Retrieval}
\label{alg:cross_modal_temporal_rrf}
\begin{algorithmic}[1]
\REQUIRE Ranked frame lists $\text{TopK}_1$, $\text{TopK}_2$, $\text{TopK}_3$; temporal window size $w_d$.
\ENSURE Final re-ranked list $\mathcal{R}$, the first query is chosen for the central and return.

\STATE Initialize a set of three-frame tuples from three Topk-results, $\mathcal{S} \gets \text{TopK}_1 \times  \text{TopK}_2 \times \text{TopK}_3$
\STATE Initialize $\mathcal{R} \gets []$
\FOR{each frame $f_1$ in $\text{TopK}_1$}
    \STATE $\mathcal{R}[f_1] \gets \underset{(f_1,f_2, f_3)\in \mathcal{S}}{\max} \Bigl(\dfrac{1}{100 + r_1}+\dfrac{1}{100 + r_2}\mathbf{1}[0<f_2-f_1<w_d]+\dfrac{1}{100 + r_3}\mathbf{1}[0<f_2-f_1, f_3-f_2<w_d]\Bigr)$
\ENDFOR
\STATE Sort $\mathcal{R}$ by descending $s_{\max}$ to obtain the final fused ranking.
\RETURN $\mathcal{R}$
\end{algorithmic}
\end{algorithm}

\subsubsection{Cross-modal Temporal Event Retieval:}
We enhance keyframe retrieval by combining semantic, object, and OCR-based queries. In a three-scene temporal sequence, one scene can use metadata-based attributes while the others use semantic text. The results from all three queries are merged and re-ranked by the Temporal Re-ranking block, as shown in \hyperref[fig:search]{Figure 3} and detailed in \hyperref[alg:cross_modal_temporal_rrf]{Algorithm 2}. Our approach selects the first scene as a pivot and subsequently retrieves the optimal sequence of frames by maximizing the accumulated re-ranking scores. We assign 10 to window size because it can cover at least three distinct scenes behind the current one, adequately evaluate about neighbor scenes and optimize the performace.

%% file: contents/UI.tex
\section{User Interface}
\begin{figure}[h]
\centering
        \includegraphics[width=0.8\linewidth]{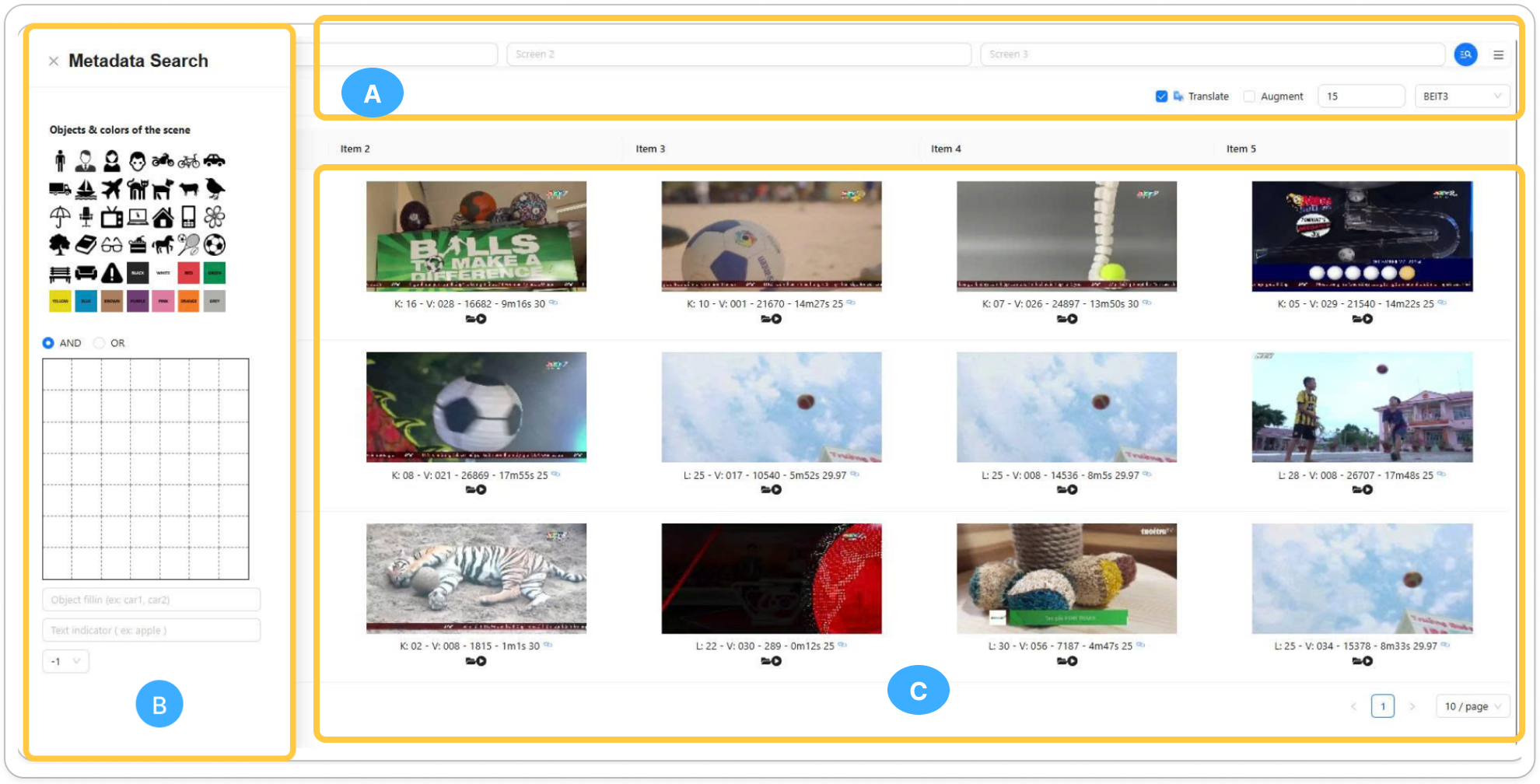}
        \caption{The user interface}
        \label{fig:ui11}
    \end{figure}
The primary objective of our system is to improve the efficiency of processing, organizing, and retrieving multimodal data within large-scale collections. See \hyperref[fig:ui11]{Figure 4} to get the overview. 
\subsection{Overall UI}
\subsubsection{Section A:}
The first section includes query input bars for entering all queries needed for the embedding-based search (CLIP and BEiT3). Next to them are buttons for running the search and opening the Advanced Search window (Section B). Below is a control panel where users can adjust search parameters such as translation, augmentation, top-k results, and model type used in Section A.
\subsubsection{Section B:}
This interface provides draggable objects that can be placed on a 7x7 dropdown grid storing detection vectors for metadata-based search. Above the grid is a radio button group to choose logical operators (AND/OR) defining relationships between search conditions. Below it are two input bars for object-tag and OCR queries, and at the bottom is a dropdown list for selecting the scene retrieved through metadata-based search, enabling cross-modal temporal event retrieval.
% \begin{figure}[ht]
%     \centering
%     \includegraphics[width=1\linewidth]{figs/pic2.png}
%     \caption{The second component is the Metadata Search interface in Section B where user can drag and drop the specified objects on a 7x7 
% grid and define all the desired search conditions. The search results from Sections A and Section B are displayed in Section C, following the same format as previously described. 
% }
%     \label{fig:ui2}
% \end{figure}
\subsubsection{Section C:}
% \begin{figure}[h]
%     \centering
%     \includegraphics[width=0.8\linewidth]{Latex-Template-for-Springer/figs/fixpic.jpg}
%     \caption{Results Display Interface. We provide a friendly UI with essential features for inspection and evaluation.}
%     \label{fig:ui2}
% \end{figure}
The interface of Section C displays the final results of the metadata-based searches from Section A and Section B on a grid. Each result shows a representative frame of the retrieved video segment with details such as video ID, timestamp, and frame index. It also provides a folder opener and a video player icon that launches a component showing 10 surrounding frames and opens the video at that exact position. A pagination bar at the bottom allows users to navigate between result pages using directional arrows, a highlighted current page indicator, and a dropdown to choose how many results to display per page (e.g., 10, 20, or 50).

\subsection{System Usage}
\subsubsection{The Impact of Query Augmentation on the Accuracy of Video Retrieval Systems:}
\begin{figure}[H]
    \centering
    \includegraphics[width=0.9\linewidth]{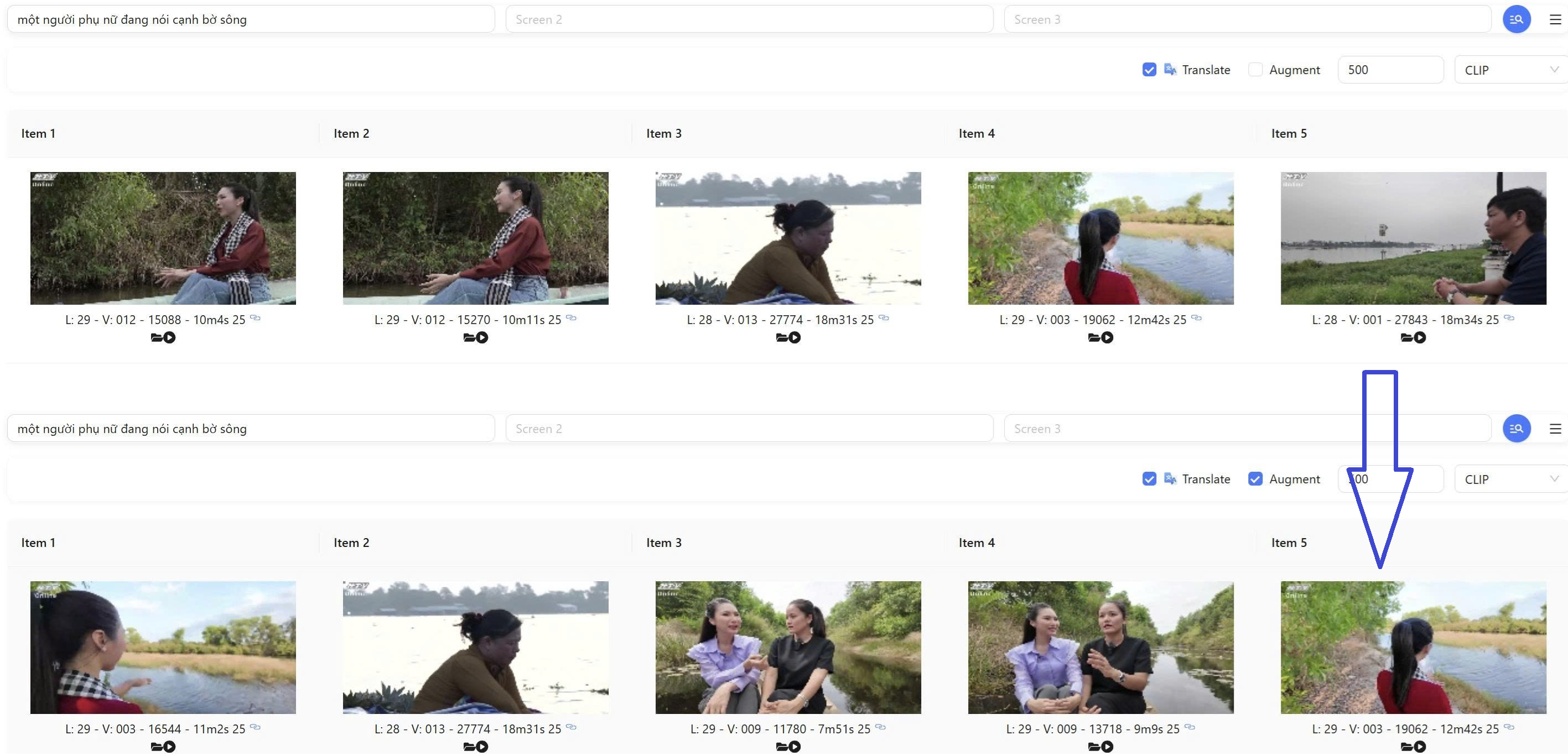}
    \caption{Retrieval results without and with query augmentation.}
    \label{fig:cs2}
\end{figure}
With the same query, if query augmentation is not applied, the model may produce incorrect results due to ambiguous wording or improper phrasing (for example, \hyperref[fig:cs2]{Figure 5}, returning an image of a man by the river instead of the woman in the upper-right corner).

When query augmentation is applied, the system refines and expands the description to better align with the model's semantic space, resulting in more accurate representations and retrieval outcomes that better match the intended target.
This case, \hyperref[fig:cs2]{Figure 5}, clearly demonstrates the impact of query augmentation in reducing noise and improving accuracy in multimodal video retrieval.

\subsubsection{Temporal and Multimodal Robustness in Video Retrieval:}
\begin{figure}[h]
    \centering
    \includegraphics[width=0.8\linewidth]{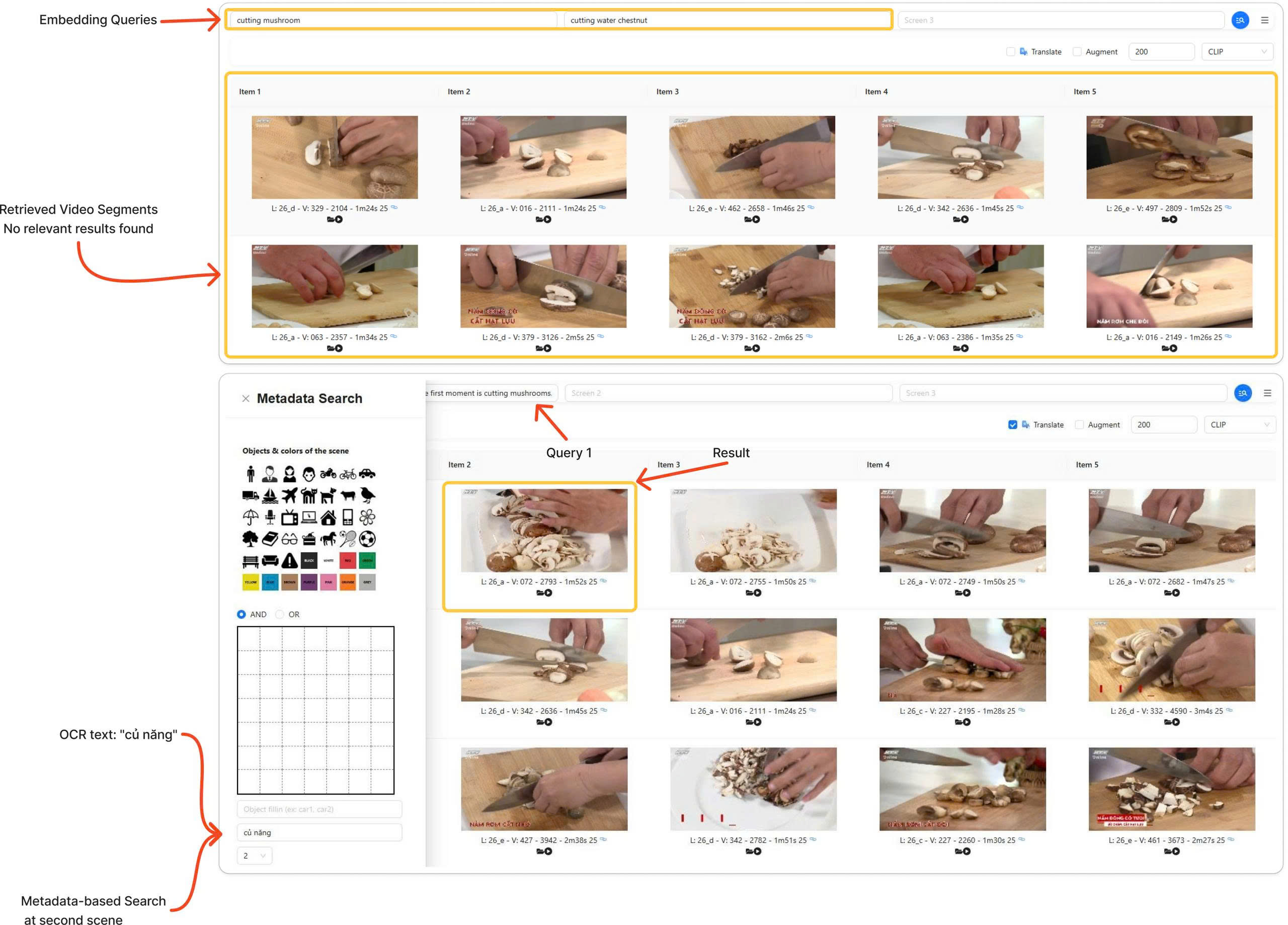}
    \caption{This user study case presents a demonstration of the effectiveness of cross-modal temporal event retrieval within a case our framework.}
    \label{fig:ui3}
\end{figure}

During the first round, there is a temporal query describing actions \textit{cutting mushrooms} and \textit{cutting water chestnut} occur sequentially. Neither the BEiT3 nor CLIP models retrieved the correct video segments, exposing a limitation in capturing contextual continuity and temporal variation between semantically related queries. This indicated that conventional text-video alignment alone was insufficient. To resolve this, the second query \textit{cutting water chestnut} was replaced with its Vietnamese equivalent \textit{cu nang} in the OCR-extracted text, which successfully retrieved the relevant frames (as the phrase \textit{cu nang} appears in the scene). See \hyperref[fig:ui3]{Figure 6} for more details.
\subsubsection{Our Final Round Results in Ho Chi Minh AI Challenge 2025}: Overral, we achieved good scores for three tasks KIS, QA and TRAKE; proving our effective and competitive retrieval system.

%% file: contents/conclusion.tex
\section{Conclusion}
This paper introduces a novel video retrieval system with diverse query modalities such as embedding-based, object and color detection, and OCR. We also use Gemini to refine the textual query, then optimize the CLIP/ BEiT3 encoder inputs and mitigate the user query ambiguity. Furthermore, to enhance the system capability in representing complex temporal queries, it allows using multiple query modalities across different scenes, thereby improving the overall generalization and effectiveness of temporal retrieval.